\documentclass{acm_proc_article-sp}
\usepackage{amsmath}
\usepackage{amssymb}

\usepackage{color}
\usepackage{dcolumn}
\usepackage{float}
\usepackage{graphicx}
\usepackage[latin9]{inputenc}
\usepackage{multirow}
\usepackage{rotating}
\usepackage{subfigure}
\usepackage{psfrag}
\usepackage{tabularx}
\usepackage[hyphens]{url}
\usepackage{wrapfig}
\usepackage{fancyvrb}

\usepackage[bookmarks, bookmarksopen, bookmarksnumbered]{hyperref}
\usepackage[all]{hypcap}
\sloppypar
\begin{document}

\title{Metadata Management in Scientific Computing}

\numberofauthors{1}

\author{
  \alignauthor
  Eric L. Seidel\\
  \affaddr{City College of New York}\\
  \email{eseidel01@ccny.cuny.edu}
}

\toappear{
  Permission to make digital or hard copies of all or part of this work
  for personal or classroom use is granted without fee provided that copies are
  not made or distributed for profit or commercial advantage and that copies bear
  this notice and the full citation of the first page. To copy otherwise, or
  republish, to post on servers or to redistribute to lists, requires prior
  specific permission and/or a fee. Copyright \copyright JOCSE, a supported
  publication of the Shodor Education Foundation Inc.
}

\maketitle
\begin{abstract}
  Complex scientific codes and the datasets they generate are in need
  of a sophisticated categorization environment that allows the
  community to store, search, and enhance metadata in an open, dynamic
  system.  Currently, data is often presented in a {\it read-only}
  format, distilled and curated by a select group of researchers. We
  envision a more open and dynamic system, where authors can publish
  their data in a {\it writeable} format, allowing users to annotate
  the datasets with their own comments and data. This would enable the
  scientific community to collaborate on a higher level than before,
  where researchers could for example annotate a published dataset
  with their citations.

  Such a system would require a complete set of {\it permissions} to
  ensure that any individual's data cannot be altered by others unless
  they specifically allow it. For this reason datasets and codes are
  generally presented read-only, to protect the author's data;
  however, this also prevents the type of social revolutions that the
  private sector has seen with Facebook and Twitter.

  In this paper, we present an alternative method of publishing codes
  and datasets, based on
  Fluidinfo\footnote{\url{http://www.fluidinfo.com}}, which is an
  openly writeable and social metadata engine. We will use the
  specific example of the Einstein Toolkit, a shared scientific code
  built using the Cactus Framework, to illustrate how the code's
  metadata may be published in writeable form via Fluidinfo.
\end{abstract}

\section{Introduction} % (fold)
\label{sec:Introduction}
Data management is quickly becoming a challenge in large scale
simulations and modeling as compute resources increase in size, and
simulations integrate with observational and experimental data. Not
only do these simulations produce increasingly large datasets, which
must then be analyzed and categorized, but the codes themselves become
more and more complex, often being developed by distributed teams. The
Cactus Computational
Toolkit\footnote{\url{http://www.cactuscode.org}} is one such
software framework, comprising over 500 software modules (known as
\emph{Thorns}), of which a subset must be compiled to produce a full
simulation stack.

The Cactus Thorns specify their public interface using the Cactus
Configuration Language (CCL), which describes the mechanics of the
thorn, but provides little semantic data. This makes it difficult to
determine which of the hundreds of thorns may be needed for a
particular simulation. There are two standard methods for dealing with
these ambiguities:

\begin{enumerate}
\item Detail the semantics of every thorn in documentation within the
  source tree. This is somewhat helpful when a user has already
  downloaded the thorn in question, but it does not help a new user
  discover useful thorns.
\item Collect documentation and use-cases for each thorn on the main
  webpage for the framework. This is much more helpful to new users in
  search of thorns, but it raises new issues. Who maintains the
  website and keeps the web-based documentation synchronized with the
  source code? Thorns are generally maintained by individual authors,
  not the community, so should all authors have write access to the
  web server? If so, how does one prevent authors from misrepresenting
  each other's codes? The end user is still presented a
  \emph{read-only} interface, meaning a user cannot easily annotate
  and recommend useful thorns to others.
\end{enumerate}

In the following sections, we will describe how Fluidinfo may be used
to annotate these datasets in a writeable manner, while preserving the
safety and integrity of the author's original data. We aim to show
that the concept of ``tagging,'' as introduced by social networking
services, is well suited to building and maintaining distributed scientific
collaborations in the computational sciences. Our approach is based on
loosely structured data, in contrast to other data formats used in
metadata and semantic web research. Section~\ref{sec:Related Work}
examines other approaches to similar problems.
Section~\ref{sec:Cactus Configuration Language} describes the
{\it Cactus Configuration Language}, which contains a substantial
amount of Thorn metadata. Section~\ref{sec:Fluidinfo} introduces
Fluidinfo, the writeable metadata engine, and its core
concepts. Section~\ref{sec:Fluidinfo+Cactus} describes specifically
our strategy for publishing the Einstein Toolkit metadata to
Fluidinfo. Section~\ref{sec:Future Work} investigates how the strategy
presented in Section~\ref{sec:Fluidinfo+Cactus} may be adapted for
publishing datasets as opposed to codes. Section~\ref{sec:Educational
  Experience} reflects on the educational value of this project, and
the Blue Waters Undergraduate Petascale Education Program that
supported it.
% section Introduction (end)

\section{Related Work}
\label{sec:Related Work}

Before discussing our approach to solving this problem, let us examine
other systems that could be used to support distributed collaboration.
\emph{RDFPeers}~\cite{Cai2004109} is a distributed RDF repository
designed to solve scalability issues faced by many centralized
metadata stores. It uses a peer-to-peer architecture to spread
metadata across many machines, and efficiently route queries to the
appropriate machine. A distributed system like RDFPeers would be a
natural fit for our problem, as it could encourage authors to maintain
the metadata pertaining to their codes and datasets alongside the
actual data; however, we feel that RDF as a data format may be
excessively complex for our purposes. We believe that a simpler format
based on social tagging, like that used by the Delicious bookmarking
service\footnote{\url{http://www.delicious.com}}, would be sufficient
for our needs. In particular, RDF is based on triples of
\emph{subjects}, \emph{predicates}, and \emph{objects}, whereas the
tagging method we describe only needs \emph{objects} and
\emph{attributes}. Clearly we could use RDF triples with a constant
predicate \texttt{hasAttribute}, but we gain little by doing so and
incur additional complexity.

% \subsection{Social Accessibility}
% \label{sec:social-accessibility}

The \emph{Social Accessibility}~\cite{Takagi:2008:SAA:1414471.1414507}
project attempts to help site-owners keep up with accessibility
standards by crowd-sourcing some of the work. It is comprised of three
pieces: (1) a browser script with which end-users may register
complaints about websites and receive patches, (2) a browser plugin to
allow volunteers to investigate accessibility issues and submit
patches, and (3) a server that stores the complaints and patches. When
an end-user visits a website, the browser script searches the server
for any applicable patches, retrieves them, and applies them to the
page. The user's browsing experience is immediately enriched by the
knowledge of the community with little effort on the user's part. This
project appears to have a similar goal to our own, enriching content
via collaborative editing, albeit applied to a different problem
domain.

\section{Cactus Configuration Language} % (fold)
\label{sec:Cactus Configuration Language}

\begin{figure}[t!]
  \centering
  \includegraphics[width=0.5\linewidth]{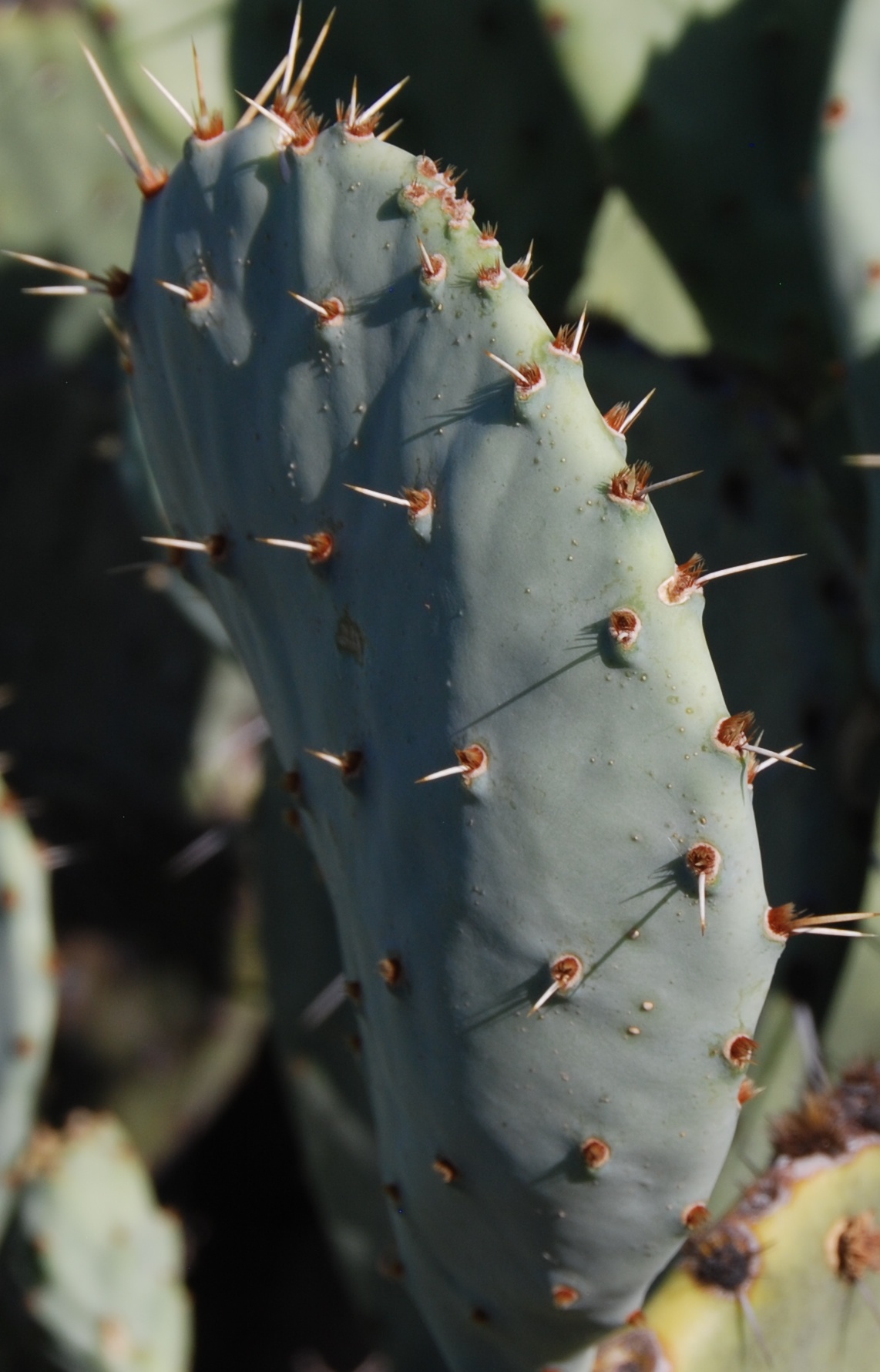}
  \label{cactusfig}
  \caption{Cactus components are called {\it thorns} and the
    integrating framework is called the {\it flesh}. The interface
    between thorns and the flesh is provided by a set of configuration
    files writing in the Cactus Configuration Language (CCL\@).}
\end{figure}

The Cactus Framework~\cite{Cactusweb,Goodale02a} is an open source,
modular, portable programming environment for HPC
computing\footnote{This section was adapted from a previous paper on
  the Cactus Configuration Language~\cite{CBHPC_CCL}.}. It was
designed and written specifically to enable scientists and engineers
to collaboratively develop and perform the large--scale simulations
needed for modern scientific discoveries across a broad range of
disciplines. Cactus is well suited for use in large, international
research collaborations.  For example, the Einstein Toolkit
Consortium~\cite{einsteintoolkitweb} is a collaboration of over 60
researchers who use Cactus for research into relativistic
astrophysics, and who maintain a core set of some 175 modules.

\subsection{Architecture}
\label{cactusarchitecture}

Cactus is a component framework. Its components are called {\it
  thorns} whereas the framework itself is called the {\it flesh}
(Figure~\ref{cactusfig}). The flesh is the core of Cactus, it
provides the APIs for thorns to communicate with each other, and
performs a number of administrative tasks at build--time and
run--time. Cactus depends on three configuration files and two
optional files provided by each thorn to direct these tasks and
provide inter--thorn APIs. These files are:
\begin{itemize}
\item{\texttt{interface.ccl}} Defines the thorn {\it interface} and
  {\it inheritance} along with variables and aliased functions.
\item{\texttt{param.ccl}} Defines parameters which can be specified in
  a Cactus parameter file and are set at the start of a Cactus run.
\item{\texttt{schedule.ccl}} Defines when and how scheduled functions
  provided by thorns should be invoked by the Cactus scheduler.
\item{\texttt{configuration.ccl} (optional)} Defines build--time
  dependencies in terms of provided and required capabilities,
  e.g. interfaces to Cactus--external libraries.
\item{\texttt{test.ccl} (optional)} Defines how to test a thorn's
  correctness via regression tests.
\end{itemize}

The flesh is responsible for parsing the configuration files at
build-time, generating source code to instantiate the different
required thorn variables, parameters and functions, as well as
checking required thorn dependencies.

At run-time the flesh parses a user provided parameter file that
defines which thorns are required and provides key-value pairs of
parameter assignments.\footnote{Note that this parameter file is
  different from the file {\texttt{param.ccl}} which is used to define
  which parameters exist, while the former is used to assign values to
  those parameters at run-time.} The flesh then activates only the
required thorns, sets the given parameters, using default values for
parameters which are not specified in the parameter file, and creates
the schedule of which functions provided by the activated thorns to
run at which time.

The Cactus flesh provides the main iteration loop for simulations
(although this can be overloaded by any thorn) but does not handle
memory allocation for variables or parallelization; this is performed
by a {\it driver} thorn. The flesh performs no computation of its own
--- this is all done by thorns. It simply orchestrates the
computations defined by the thorns.

\begin{figure}[t!]
  \centering
  \includegraphics[width=\linewidth]{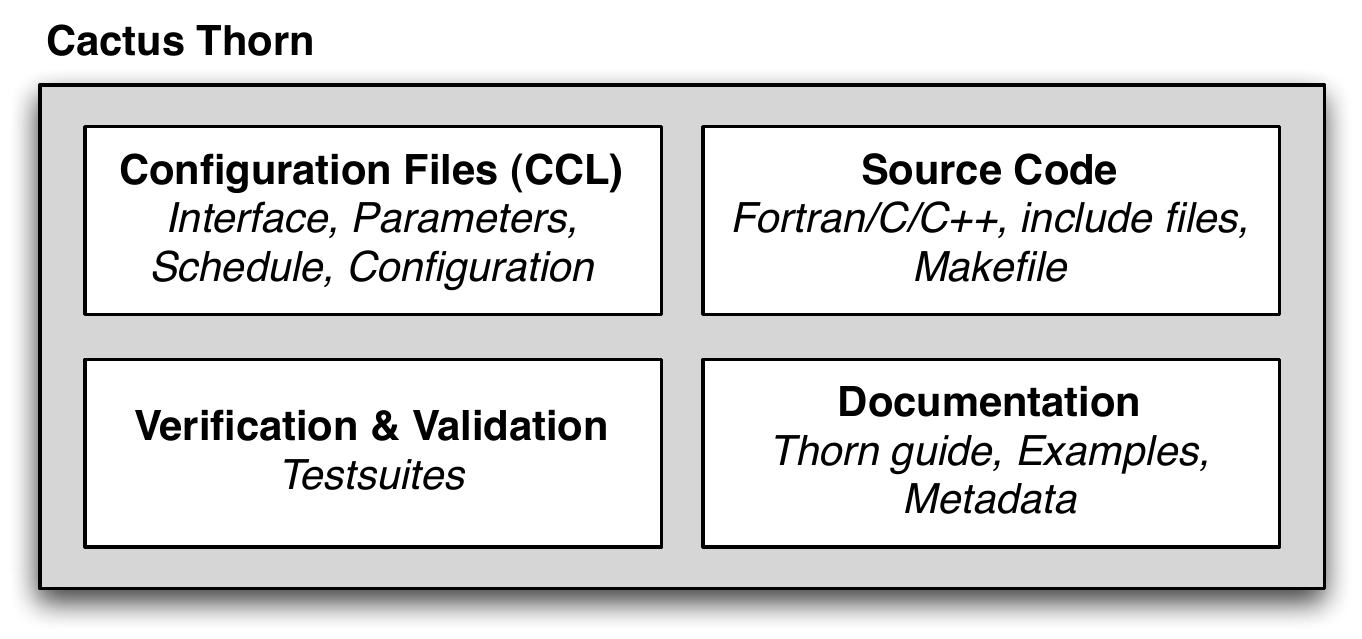}
  \label{thornfig}
  \caption{Cactus thorns are comprised of source code, documentation,
    test--suites for regression testing, along with a set of
    configuration files written in the Cactus Configuration Language
    (CCL) which define the interface with other thorns and the Cactus
    flesh.}
\end{figure}

The thorns are the basic modules of Cactus. They are largely
independent of each other and communicate via calls to the Flesh
API\@. Thorns are collected into logical groupings called
\emph{arrangements}. This is not strictly required, but strongly
recommended to aid with their organization. An important concept is
that of an \emph{interface}. Thorns do not define relationships with
other specific thorns, nor do they communicate directly with other
thorns. Instead they define relationships with an interface, which
may be provided by multiple thorns. This distinction exists so that
thorns providing the same interface may be interchanged
without affecting any other thorns. Interfaces in Cactus are fairly
similar to abstract classes in Java or virtual base classes in C++,
with the important distinction that in Cactus the interface is not
explicitly defined anywhere outside of the thorn.

This ability to choose among multiple thorns providing the same
interface is important for introducing new capabilities in Cactus with
minimal changes to other thorns, so that different research groups can
implement their own particular solver for some problem, yet still take
advantage of the large amount of community thorns. For example, the
original driver thorn for Cactus which handles domain decomposition
and message passing is a unigrid driver called {\tt PUGH}. More
recently, a driver thorn which implements adaptive mesh refinement
(AMR) was developed called \texttt{Carpet}~\cite{Schnetter-etal-03b,
  Schnetter06a, ES-carpetweb}. Carpet makes it possible for
simulations to run with multiple levels of mesh refinement, which can
be used to achieve great accuracy compared to unigrid
simulations. Both \texttt{PUGH} and \texttt{Carpet} provide the
interface \texttt{driver} and application thorns can relatively
straightforwardly migrate from unigrid to using the advanced AMR
thorn.

Thorns providing the same interface may also be compiled together in
the same executable, with the user choosing in the parameter file, at
run-time, which implementation to use.  This allows users to switch
among various thorns without having to recompile Cactus.

Thorns include a \texttt{doc} directory which provides the
documentation for the thorn in \LaTeX\ format. This allows users to
build one single reference guide to all thorns via a simple command.

\subsection{Tools}
As a distributed software framework, Cactus can make use of some
additional tools to assemble the code and manage the simulations.
Oftentimes each arrangement of thorns resides in its own source
control repository, as they are mostly independent of each other. This
leads to a retrieval process that would quickly become unmanageable
for end-users (for example the Einstein Toolkit is comprised of 135
thorns). To facilitate this process we use a \emph{thornlist} written
using the Component Retrieval Language~\cite{TG_CRL}, which allows the
maintainers of a distributed framework to distribute a single file
containing the URLs of the components and the desired directory
structure. This file can then be processed by a program such as our
own \texttt{GetComponents} script, and the entire retrieval process
becomes automated.

In addition to the complex retrieval process, compiling Cactus and
managing simulations can be a difficult task, especially for new
users. There are a large number of options that may be required for a
successful compilation, and these will vary across
architectures. To assist with this process a tool called the
\emph{Simulation Factory}~\cite{ES-simfactoryweb, CBHPC_SimFactory}
was developed. Simulation Factory provides a central means of control
for managing access to different resources, configuring and building
the Cactus codebase, and also managing the simulations created using
Cactus. Simulation Factory uses a database known as the \emph{Machine
  Database}, which allows Simulation Factory to be resource agnostic,
allowing it to run consistently across any pre-configured HPC
resource.
% section Cactus Configuration Language (end)

\section{Fluidinfo} % (fold)
\label{sec:Fluidinfo}
\begin{figure*}[ht!]
  \centerline{\includegraphics[width=\linewidth]{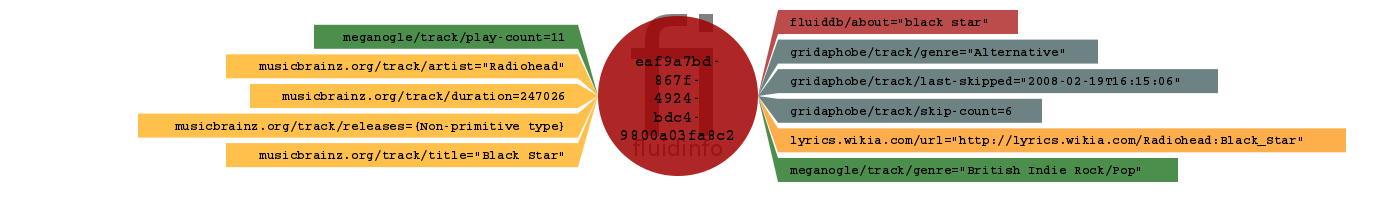}}
  \caption{\label{fig:black-star}\small{Visual representation of the
      Fluidinfo object for the song ``Black Star'' by Radiohead. Note
      the combination of tags from a variety of users, with primitive
      and opaque values.}}
\end{figure*}
Fluidinfo is an openly writeable datastore, whose goal is to extend
collaborative tagging to all forms of data. Designed around the
metaphor of post-it notes, it is a collection of {\it objects} and
{\it tags} at its core, with a complete set of {\it permissions} to
give users full control over their data. Fluidinfo is developed and
hosted by Fluidinfo Inc., a start-up company. This section will give a
brief overview of the basic concepts of Fluidinfo; a more detailed
discussion may be found in the official
documentation~\cite{fluid-docs}.

\subsection{Objects}
\label{sec:Objects}
One of the core concepts of Fluidinfo is that objects are completely
anonymous, having no owner and no inherent meaning. Objects exist
solely as a container for tags, which define their semantics.

\subsection{Tags}
\label{sub:Tags}
Tags have owners and permissions, so while anyone can tag an object,
tags may be read-only, read-write, or completely invisible to the
outside world. When a tag is placed on an object, it may contain any
value, and the type of value need not be consistent between
tag-instances (although in practice this would be a good
idea). Fluidinfo does, however distinguish between so-called {\it
  primitive} and {\it opaque} tag-values.
\begin{description}
\item[Primitive tag-values] are a subset of the standard types found
  in many programming languages: integers, floating-point numbers,
  booleans, strings, the {\it null} value, and {\it sets of
    strings}. Note that arrays, or sets of anything other than strings
  are considered opaque values. Primitive values are useful because
  Fluidinfo allows indexing of these values, permitting more complex
  and specific querying of tags with primitive values.

\item[Opaque tag-values] include any type of value that is not
  considered primitive. This includes JSON arrays or objects, binary
  data, anything that can be assigned a MIME-type. Opaque values are
  not indexed, and therefore users cannot search based on the contents
  of opaque tags, merely their presence.
\end{description}

\subsubsection{About Tag} % (fold)
\label{ssub:About Tag}
If objects are anonymous and an instance of a tag may contain any
value independent of the other instances, one may wonder how to
identify a specific object. Fluidinfo allows objects to be uniquely
identified by a {\it UUID} (Universally Unique
Identifier\footnote{\url{http://en.wikipedia.org/wiki/UUID}}) and the
so-called {\it about-tag}. The about-tag, {\tt fluiddb/about}, is a
unique, immutable tag that may optionally be provided when creating an
object. This allows for an object to be given some basic semantic
value without adding any user tags to it, which can be useful in
establishing tagging conventions.
% subsubsection About Tag (end)

\subsection{Namespaces}
\label{sub:Namespaces}
Tags can be grouped together in {\it Namespaces}. All of a user's tags
will live inside the user's top-level namespace to avoid conflicts
with other users' tags, but sub-namespaces can be used to logically
group tags. As an example, suppose the Fluidinfo user {\tt eric}
created a {\tt rating} tag in his top-level namespace, the qualified
name of that tag would be {\tt eric/rating}. If we look back at
Section~\ref{ssub:About Tag}, we can surmise that there is actually
nothing special about the about-tag, it is simply a tag belonging to
the {\tt fluiddb} user, who is guaranteed to never change the value.

\subsection{Permissions} % (fold)
\label{sub:Permissions}
The core mechanic that allows Fluidinfo to be flexible is its {\it
  permissions} system. Each namespace and tag has an explicit set of
permissions, describing exactly how users may interact with the item
in question. This affords users fine-grained control over their data.
They can publish it in read-only, read-write, or write-only form, or
even transfer entire control of a namespace/tag to another
user\footnote{For a more detailed and complete list of the allowed
  permissions, visit
  \url{http://doc.fluidinfo.com/fluidDB/permissions.html}}. As an
example of how these permissions can be used, let us examine how
Fluidinfo creates new users. There is a tag, {\tt
  fluiddb/users/username}, placed on the object representing a user,
that tells Fluidinfo that such a user exists. The {\tt fluidinfo.com}
user has {\it create} permissions for this tag, so when a new user
signs up on \url{http://fluidinfo.com}, the {\tt fluidinfo.com} user
creates a new object and adds the {\tt fluiddb/users/username} tag to
it, signifying that a new user has been created.
% subsection Permissions (end)

\subsection{Fluidinfo Query Language} % (fold)
\label{sub:Fluidinfo Query Language}
Fluidinfo includes a simple query language to allow users to search
the datastore for specific tags and tag-values. There are five basic
types of queries in Fluidinfo's query language.
\begin{description}
\item[Presence] queries are the simplest type. They check only for the
  presence of a tag on an object, and are written as {\tt has <tag>}.
\item[Numeric] queries search for tags that have a specific value
  using the standard mathematical equality operators, and are written
  as {\tt <tag> (=,<,>,etc.) <value>}.
\item[Textual] queries attempt to match the query text against the
  text contents of a tag, and are written as {\tt <tag> matches
    <text>}.
\item[Set contents] queries check for the tags that contain the given
  string. Note the difference between set contents and textual
  queries: set contents apply to tags containing a {\it set of
    strings} while textual queries apply to tags containing a single
  string. Set contents queries are written as {\tt <tag> contains
    <string>}.
\item[Logical] queries combine the above types using the {\tt (}, {\tt
    )}, {\tt and}, {\tt or}, and {\tt except} operators. This allows
  arbitrarily complex queries, such as\\{\tt (has eric/seen and
    (eric/rating > 4 or john/rating > 8)) except imdb.com/rating <~5}.
\end{description}
% subsection Fluidinfo Query Language (end)
% section Fluidinfo (end)

\section{Writeable Metadata Engine for Cactus Components}
\label{sec:Fluidinfo+Cactus}
In this section we will describe the desired capabilities for handling
metadata for simulation codes, such as the ability to support open
data objects and metadata which can then be provided by any user,
promoting community driven standards and enabling innovation. Such a
system would allow researchers to annotate the codes with their
opinions, experiences, or tips while preserving the integrity of the
original data. Social networks have already solved a subset of this
problem, but there is no equivalent system in use by the scientific
community.

Foursquare is a location-aware social networking site. Users publish
their presence at a physical location, e.g. a metro station, a
restaurant, a school, and can add photos or tips for others. If the
physical location does not exist, users can add their own selecting
basic metadata to describe the site. Social features include the
ability to see where your friends are and have been, and to read the
tips left by others.

Learning from this flexible model we envisage similar tools for data
that will encourage academic data to break free from the current
constraints of rigid schema, proprietary and controlled databases and
lack of social networking tools. The general scenario we envisage is
described below, here for software components, although a similar
methodology will work for general data sets.

\begin{enumerate}
\item Software components (e.g. Cactus Thorns) are added to Fluidinfo
  in the same manner as foursquare locations. Basic tags could for
  example be based on the Dublin Core~\cite{dublincoreweb}, with
  fields for authors, software location, etc. These tags can only be
  edited by the original user unless specified otherwise. An object
  would be created in Fluidinfo for each software component, we
  suggest an about-tag convention of {\tt CCTK:<arrangement>/<thorn>};
  however, this is strictly optional as the thorns would also be
  identified by their tags. There could also be multiple objects for
  each thorn since they could be added by people other than the
  original authors.

\item Trusted experts or consortia can then tag thorns to provide a
  quality ranking, associate datasets generated by the thorn, or warn
  new users of an existing bug. For example, a maintainer for the
  Einstein Toolkit would tag Cactus thorns with the release for which
  they have been tested and verified. Users can then search for
  software which has been ratified by the Einstein Toolkit Consortium,
  or they could search for software that has been recommended by a
  trusted colleague.

\item A graduate student is working on a research project to develop a
  new ontology for scientific computing. She can easily add tags
  representing this ontology to the Cactus thorns, where the user
  community can test out her work without necessitating new servers,
  or without her having write access to the basic thorn tags.
\end{enumerate}
\begin{figure*}
  \centerline{\includegraphics[width=\linewidth]{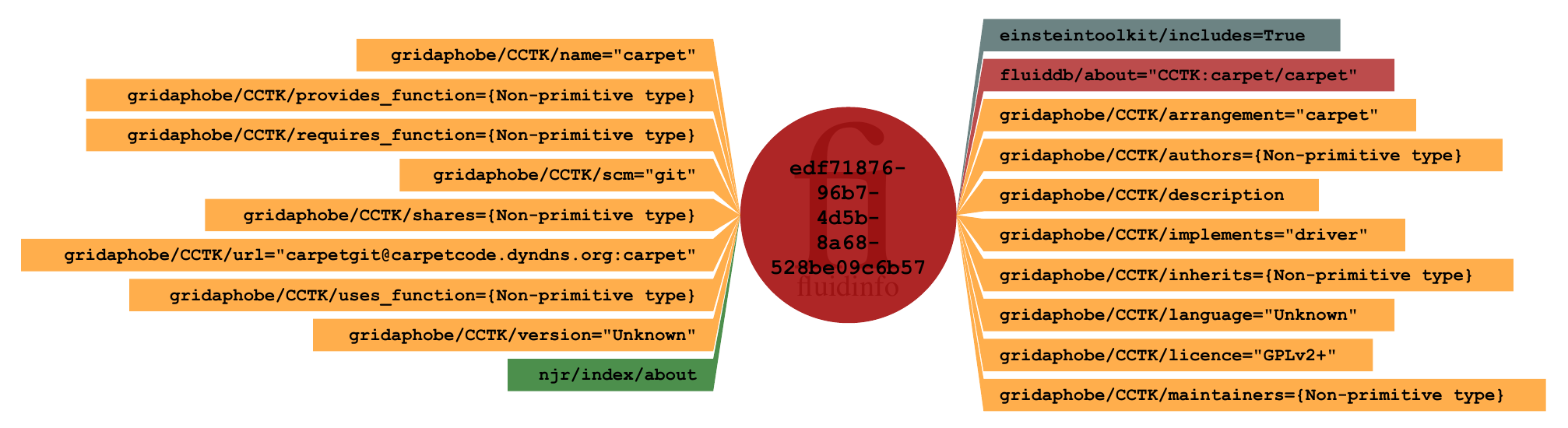}}
  \caption{\label{fig:carpet}\small{Visual representation of the
      Fluidinfo object for the Carpet module in the Einstein
      Toolkit.}}
\end{figure*}

We implemented a prototype of such a system for the Cactus Thorns,
with a web front-end written in Python~\cite{pythonweb}. The initial
set of metadata we extracted from each thorn came from the
configuration files and the Readme, representing a subset of the
functional and bibliographical metadata contained in each thorn, as
seen in Table~\ref{tab:tags}.
\begin{table*}
  \centering
  \begin{tabular}{|l|p{100mm}|}
    \hline
    {\bf Fully-qualified Tag} & {\bf Description} \\ \hline \hline
    {\tt gridaphobe/CCTK/arrangement} & The {\it arrangement} the thorn belongs to.\\ \hline
    {\tt gridaphobe/CCTK/authors} & A list of all authors of the thorn.\\ \hline
    {\tt gridaphobe/CCTK/description} & The description of the thorn as found in the Readme.\\ \hline
    {\tt gridaphobe/CCTK/implements} & A list of {\it interfaces} the thorn implements.\\ \hline
    {\tt gridaphobe/CCTK/inherits} & The thorn (if any) inherited from.\\ \hline
    {\tt gridaphobe/CCTK/name} & The name of the thorn.\\ \hline
    {\tt gridaphobe/CCTK/scm} & The version control system used for the thorn's source code.\\ \hline
    {\tt gridaphobe/CCTK/url} & The URL where the thorn's source code is located.\\ \hline
  \end{tabular}
  \caption{\label{tab:tags}A sample of the tags used to describe Cactus thorns in Fluidinfo.
    The tag names are fully-qualified and assume the current user's name is
    {\tt gridaphobe}.}
\end{table*}
These tags are added automatically by a Python script that parses the
configuration and Readme files of a thorn. The intent is for thorn
authors to run this script on their thorns, immediately populating
Fluidinfo with a set of Cactus metadata. Once the basic set of
metadata has been imported, we can begin to enhance the existing data
by adding other relevant tags to the objects representing thorns.

The Einstein Toolkit is a small subset of all Cactus thorns, and
thorns may be imcompatible with each other, e.g. if they implement the
same interface. Therefore it would be useful for users to know if any
given thorn is part of the Einstein Toolkit; we can implement this
quite naturally by creating an {\tt einsteintoolkit.org}
user\footnote{Fluidinfo only allows the owner of a domain to create
  the user for that domain, so domain users can be more readily
  trusted.}, which will tag all thorns in the toolkit with an {\tt
  einsteintoolkit.org/includes} tag with the value set to {\tt
  True}. Figure~\ref{fig:carpet} illustrates what the resulting
Fluidinfo object might look like.

\begin{figure}
  \centering
  \includegraphics[width=\linewidth]{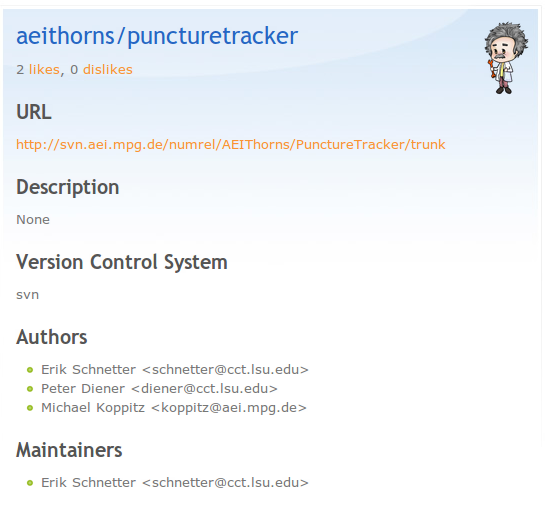}
  \caption{\label{fig:cactuse}A prototype of a web application that
    dynamically displays thorn metadata based on the tags stored in
    Fluidinfo. The Einstein logo in the top-right corner indicates
    that this thorn is part of the Einstein Toolkit.}
\end{figure}

Using this tag structure we created a simple web application, running
on Google's AppEngine platform, to dynamically retrieve the objects
representing the Einstein Toolkit, and insert the values into an HTML
template for easy viewing of the thorn
metadata. Figure~\ref{fig:cactuse} shows a sample page from this web
application.

With these two sources of data we can already perform useful queries
on the Cactus metadata. Cactus uses a tool called {\it
  GetComponents}~\cite{TG_CRL} to automate the process of retrieving
many thorns from different locations. To accomplish this,
GetComponents essentially needs three pieces of information:
\begin{enumerate}
\item Where the thorn is located (URL).
\item How to retrieve the thorn (version control system).
\item Where to place the thorn on the local filesystem.
\end{enumerate}
All of this data is contained in the Fluidinfo tags posted by the
Python script\footnote{Cactus has a convention of placing thorns
  inside an {\tt arrangements} directory with the structure {\tt
    arrangements/<arrangement>/<thorn>}.}!  So if we wanted to retrieve
the Einstein Toolkit, we could dynamically generate a file in the CRL
format GetComponents uses by querying Fluidinfo for all objects that
have {\tt einsteintoolkit.org/includes = True}, retrieving the tags
\begin{itemize}
\item {\tt gridaphobe/CCTK/arrangement}
\item {\tt gridaphobe/CCTK/name}
\item {\tt gridaphobe/CCTK/url}
\item {\tt gridaphobe/CCTK/scm}
\end{itemize}
The returned data could then be reformatted into a CRL file, and
GetComponents invoked to automatically retrieve the requested
thorns\footnote{There are some issues not covered by this example,
  e.g. the directory structure of different git repositories, but none
  that could not be resolved by adding a few extra tags}.

This is already a significant improvement over the current system of
creating and distributing a thornlist, which is both tedious and
error-prone, but we can go further and solve a problem that was
previously unsolvable. The Einstein Toolkit thorns can all be compiled
together; however, they are not all needed to run individual
simulations. Researchers will generally only compile a subset of the
Einstein Toolkit, including just the thorns needed to model their
particular system. In this case downloading the entire Einstein
Toolkit is superfluous, we would like to simply download the thorns
that we {\it actually need}. Using the thorn configuration files, we
can construct a list of the thorns we will need to download in order
to use a specific {\it base set} of thorns, providing initial data,
drivers, and other components of a simulation~\cite{CBHPC_CCL}. We
can then dynamically retrieve the tags mentioned above for only this
subset of thorns, and provide GetComponents with a much smaller list
of thorns to download. This also has the benefit of isolating the code
in the source tree of any simulation to only that which is necessary.

If we wanted to implement a system like this on our own, we would have
to setup a new webserver and database, define a schema to contain the
data, create a REST API, and then assign someone to maintain the
database and server. If we additionally wanted the system to be
writeable (or at least have individual thorns managed by their
authors), we would then have to implement an authentication system as
well, and our data would still be limited to some pre-defined schema.
Fluidinfo allows others to add to our data, and we can {\it choose}
whether to ignore it or to begin incorporating pieces into our
applications.
% section Fluidinfo+Cactus (end)

\section{Future Work} % (fold)
\label{sec:Future Work}
In the previous section we saw how to use Fluidinfo to store the
metadata of Cactus thorns in a {\it writeable} format, add tags to
those thorns from a different source, and then use tags from both
sources to solve a problem that previously could not be solved without
setting up our own web server. We did, however, ignore one issue; the
example only dealt with thorns uploaded by one user, whereas the
Einstein Toolkit is comprised of thorns written by many different
authors. Suppose we don't know who all of the authors are, how will we
know which tags to retrieve? For example, the Carpet thorns are
written by Dr. Erik Schnetter, but unless we know his Fluidinfo
username, we won't know how to retrieve his tags. Fluidinfo does not
currently support wildcards in the list of tags to return, so we {\it
  must} explicitly list the tags we want. So how can we best adapt our
solution to the actual problem? There are two possible solutions:
\begin{enumerate}
\item Instead of using tags in the author's namespace, we could take
  advantage of Fluidinfo's permissions system to give all authors
  write permission to tags in a {\tt cactuscode.org/CCTK}
  namespace. This way we would always retrieve tags from the trusted
  domain user. This solution detracts from the personalization of
  Fluidinfo though, since the tags are coming from a domain user
  instead of the author himself. In a sense this represents how we
  might solve the metadata problem on our own, but with the extra
  downside that we can no longer prevent authors from modifying each
  other's tags! Fluidinfo does not allow separate permissions per
  tag-instance, and this would become far too complex to manage
  regardless.

\item Create a {\tt cactuscode.org/author} tag that would be applied
  to the objects representing the {\it users} in Fluidinfo who are
  authors of Cactus thorns. This way we can query Fluidinfo for the
  objects with the tag, and ask it to return the {\tt
    fluiddb/users/username} tag, giving us a list of all Fluidinfo
  users who are also Cactus authors. Then we can proceed with the
  process described in Section~\ref{sec:Fluidinfo+Cactus}. This
  solution has several advantages: (1) authors cannot modify each
  other's tags without explicit permission, (2) in the event of a {\it
    tag collision} (where more than one author has tagged a thorn) we
  can apply some filtering condition based on the thorn's own author
  list to determine which tags are most authoritative, (3) we are
  actually adding more data to the ecosystem by {\it tagging} the
  users as Cactus authors.
\end{enumerate}

\subsection{Other Datasets} % (fold)
\label{sub:Other Datasets}
Supercomputers are generating massive amounts of data on a daily
basis, data which must be stored efficiently and then classified so
that it can be referred to and even cited. Our strategy in
Section~\ref{sec:Fluidinfo+Cactus} can easily be adapted to solve this
problem. Suppose we run a simulation of two colliding neutron stars
and store the resulting dataset. We can now create an object in
Fluidinfo to represent this simulation, and tag it with the machine
used, number of cores, initial values, duration, and any number of
other relevant statistics about both the simulation and the
output. Then a PhD student uses our dataset in her thesis; she can tag
the dataset in Fluidinfo with a {\tt <student>/cited} tag whose value
would be a list of all papers in which she cited our dataset (likely
using DOIs). If she is consistent in tagging the datasets she has
cited, we could perform interesting queries using Fluidinfo, i.e. we
could quickly determine which supercomputers had contributed most to
her work. Other researchers might tag the datasets with specific
situations where they proved useful, or perhaps related datasets. With
a writeable, schemaless system, the datasets may be augmented in any
fashion deemed suitable by users. This allows for use-cases the
original publisher could not have conceived of to arise organically.

It is becoming clear that citing datasets produced by simulations will
be essential for continued scientific progress, one need look no
further than the NSF's Computational and Data-Enabled Science and
Engineering\footnote{\url{http://www.nsf.gov/mps/cds-e/}}
program. Ball and Duke have raised some important questions that will
have to be answered for data citation to become
widespread~\cite{Ball2011}. We would like to address the question of how
the metadata can be stored in a manner accessible both to humans and
automated scripts. By storing the metadata in a shared datastore like
Fluidinfo, it is immediately available for consumption by scripts, and
by extension easily converted into a human-readable page as we have
demonstrated in this paper. We also gain the advantage of not being
tied to any schema, allowing us to freely add more metadata whenever
necessary. Finally, the writable nature of Fluidinfo removes the
author's responsibility of linking to all papers that have cited the
dataset. The author of a paper can simply tag the dataset in Fluidinfo!
% subsection Other Datasets (end)
% section Future Work (end)

\section{Conclusion}
\label{sec:Conclusion}
Scientific research is increasingly dependent on the simulation of
complex processes and, by extension, on the ability to organize,
search, and refer to the datasets generated by simulations. We propose
using writable metadata to distribute and maintain scientific
metadata, and have shown one possible method of implementing such a
system. More work will be required to investigate alternative systems,
schemas, and interfaces, as well as to determine what would be an
optimal solution. We hope that the scientific community will take this
opportunity to start a conversation about how to manage the large
amounts of data currently being generated by our research on a daily
basis.

\section{Educational Experience} % (fold)
\label{sec:Educational Experience}
The research presented in this paper was performed as part of a
year-long internship sponsored by the Shodor Educational
Foundation\footnote{\url{http://www.shodor.org}}. The program began
with a two-week intensive introduction to HPC, covering
parallelization issues, N-Body problems, MPI, and other computational
science topics. Following the introductory session, the interns split
up to work with individual mentors for the rest of the year. While not
strictly related to Computational Science, the research presented in
this paper was strongly supported and enhanced by the Blue Waters
Petascale Internship, especially the focus on solving {\it real}
problems.

\section*{Acknowledgments} % (fold)
\label{sec:Acknowledgments}
This work was supported by the Blue Waters Undergraduate Petascale
Education Program, as well as Fluidinfo, Inc. The initial work
relating to the Cactus Configuration Language was supported by NSF REU
program (\#1005165). We would like to thank Steven Brandt, Frank
L{\"o}ffler, and Erik Schnetter for their mentorship in the Cactus
group, and Gabrielle Allen for suggesting the use of Fluidinfo for
storing the thorn metadata.  We acknowledge Nicholas J. Radcliffe, who
created \url{http://abouttag.com} to generate visuals of Fluidinfo
objects. Eric Seidel is currently working for Fluidinfo part-time
after a summer internship.
% section Acknowledgments (end)

\bibliographystyle{amsplain-url}
\bibliography{CCL,publications-schnetter,references,bwupep}

\end{document}